\documentclass[a4paper,10pt]{article}
\usepackage{graphicx}
\usepackage{amsmath}
\usepackage[utf8]{inputenc}
\usepackage{hyperref}
\usepackage{indentfirst}
\usepackage{array}
\usepackage{float}
\usepackage{subfig}
\newcolumntype{L}{>{\centering\arraybackslash}m{3cm}}
\begin{document}
 
\begin{center}
\mbox{}

{\large{\bfseries  {Close Friends, Popular Peers, Team formation and Leadership in Group Projects }}} 
\\[8pt]

\textbf{Gowrishankar G, Deekshajyothi S}
\\[8pt]

*E-mail: {\tt gowrishankar.is18@bmsce.ac.in, deekshajyothi.is18@bmsce.ac.in}

\date{}
\end{center}
\vspace{0.2cm}

\begin{abstract}
The paper discusses diverse interconnected relationships formed within a seemingly unrelated group of students and a conceptually different problem statement. This study uses a Social Network Analysis (SNA) to analyze and map the connections among the individuals. SNA facilitates understanding the psychology of a group of people while performing a certain task. This could help predict further patterns in which a similar group might perform these tasks.  As discussed in this paper, the analysis of selection of teammates for a project indirectly implies friendships in a department and can predict new friendships that could result from these. The data collected can be represented as a Social Network using various analysis tools such as Gephi and NodeXL. The result of the analysis determines the popular peer among the students and elucidates the reason behind it. Also, various other Inferences such as the close friends list, Social Influence of a node, and more were deduced.
\end{abstract}
\textbf{} {\bf Keywords:} SNA; Social Networks; Directed graph; In-degree; Out-degree; Graphs; Python; Team formation\\

\section{Introduction}
Studying the relationship patterns that arise when students choose their teammates for a project is interesting, as it unfolds a new spectrum of collective behavior. Understanding how the relationships in terms of academics and friendships are formed in undergraduate classrooms shows an increased impact on the learning outcomes of a student. This paper discusses a case study on Social Network Analysis, carried out at Information Science and Engineering department, B.M.S.College of Engineering (BMSCE). A survey was conducted to know the patterns of how students choose their project teammates. The goal was to analyze the reasons behind their selections. Students are involved in many group projects as part of various course assessments, namely the Alternate Assessment Tool (AAT). 
\newline
The academic institute focuses on quality assurance and improvement in the curriculum adhering to the norms and standards specified by the National Board of Accreditation (NBA). The program’s outcomes (PO) defined by the institute are PO9, PO10, PO11, PO12 specified by NBA directly maps to the project work and team performance. 
\begin{itemize}
 \item
 	PO9: Individual and teamwork
 \item
    PO10: Communication
 \item    
    PO11: Project management and finance
 \item
    PO12: Lifelong learning
\end{itemize}
All four higher-order POs contribute to team formation, individual performance and team performance, and communication skills that cater to lifelong learning principles. Group projects intend to inculcate good team-building, enhancement of communication and leadership skills. As varied personalities, thinking patterns, and problem-solving skills are different, such activities groom students to be ready for the corporate world. 
Students need to form groups for various courses across their academics. It is fascinating to see the changes in their psychological behavior while their choices in forming teams and study partners vary over the study years. Various questions such as - Are students choosing only their friends as teammates? Do the project groups vary over different semesters of their studies? Are they open to new opportunities and friendships? Are selections based on the subject knowledge of a student? Do students connect linguistically and tend to form groups? - can be answered through a detailed and continuous analysis

\section {Reference Section}
The study in Rahman et al \cite{oloritun2013identifying}, shows that close friends' interactions build a strong connection between them for a longer time. Various time zones are identified when close friends interact with each other. Indirectly it implies, the more time spent among two individuals, the stronger their connection is. Furthermore, Sepehr in \cite{saeidibonab2017homophily} talks about the homophily approach. Homophily enunciates that there is a higher chance of contact between similar people than with dissimilar people. Authors extrapolate the group formation trends considering various factors such as gender, religion, political and nationality. Group projects are a pivotal part of any curriculum that paves the way to a student’s career path. Students can understand how work can be equally distributed among teammates and complete tasks faster. This concept of collaborative, agile learning is discussed by the authors Ingrid et al. \cite{noguera2018collaborative}, mainly focusing on the online environment. Daniel Z. Grunspan et al. \cite{grunspan2014understanding}, students study partners and their pattern of interactions during the  examinations is observed. A collective decision is made based on attributes like class grades, lectures attended, sufficed with demographic information. Binary ties and valued ties represent the study partner and time spent with each other for study respectively.
\cite{pociask2017does} self designed, self and random team formations. Table 1 The instructor-designed teams (Designed) were formed by placing students into teams based on their responses to a personality survey, their year in college, gender, and their major. 
The instructor's main intention was to maximize the diversity  among the teammates.
Student-formed teams (Self) were allowed to assemble in class as the students preferred. 
Random teams formed were accomplished by the use of group-forming software in the course learning management system, Moodle.
The diversity score reflects the combination of three separate measures of diversity: student
personality, class year, and gender. Each measure in the diversity measure had a maximum of 1, thus when combined, the maximum possible diversity score was 3. In the study conducted by Mohammad Tazli Azizan, 
Nurhayati Mellon, Raihan Mahirah Ramli and Sugiar Yusup in \cite{azizan2018improving} have used a teaching strategy named cooperative learning. They conducted a study on what the students experienced and the challenges they faced during group projects.

\section{Methodology}
A form was circulated to collect data related to the choice of teammates amongst 6th-semester students in the department. Agneessens et al. \cite{agneessens2022collecting} discusses various strategies and elements to be considered during a survey in any organization. Their work provides an overview of the significant steps involved while collecting data from the peers perspective. Also key interactions within a team can be captured to know the structural foci. Based on the goal and requirements of the project, a course teacher decides the team size. The size varies from 2 to 3 per team. In our dataset, students were asked to enter the University Seat Number (USN) of the teammates they would wish to do a project for a 2-Member team and USNs of teammates for a 3-Member team.  Each department has three sections, each consisting of roughly 60 students. Some course teachers approve choosing teammates across sections, while some teachers keep it specific to the section. Therefore, students were asked if they wished to choose teammates from different sections within the department and, if yes, to enter their USN. Dataset collection leads to a directed network of nodes (students), where friendships are analyzed. 

\subsection{Pre-processing the Data}
The original dataset of responses had basic information with few errors. It had to be cleaned and preprocessed before further analysis. A python script was used to retrieve only the USN and Name of the respondents from the raw dataset and was written into another CSV file which would serve as the ‘nodes’ dataset for graphical analysis in Gephi. Any duplicates or outliers from the typical USN were removed. ‘Edges’ dataset was built, where the ‘Source’ consisted of respondent’s USN and the ‘Target’ comprised of the teammates’ USNs that the respondent selected. Hence this will result in a directed graph. The student chosen as part of the 2-Member team and a student from a different section were assigned a weight of 2 as the single person chosen first represents more closeness. In contrast, the others selected as part of the 3-Member team were assigned a weight of 3 as they were the second choice and part of a bigger group. 

\subsection{Discovering Group Dynamics through Graphs Topologies}
The resulting nodes and edges datasets were fed into the Gephi tool. Various metrics such as Average degree, Average weighted degree, Modularity, HITS, Page Rank, Network diameter, Average Clustering Coefficient were estimated through the Data laboratory. As shown in figure 1, nodes are sized from smallest to largest based on the indegree. Nodes are color-coded based on the modularity class they belong to. A modularity class consists of nodes that have similar shared interests and mutual friends.  By using modularity, the strength of the network was measured and divided into clusters. The Social Networks framed from the selection of team members can be visualized in various layout options provided, like Fruchterman Reingold in figure 1 and Force Atlas in figure 2.
As explained in \cite{gibbs2017investigating}, the types of teams also plays a crucial role in determining the group dynamics. There are few configurations like long-term, short-term groups that have an impact on the group functions. 

\begin{figure}[H]
  \centering
  \includegraphics[width=0.7\textwidth]{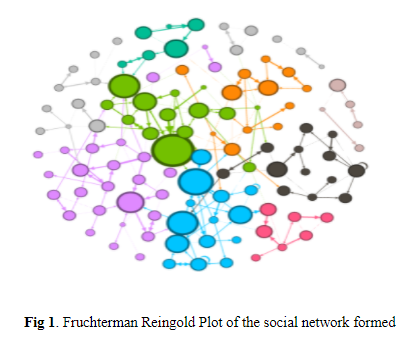}
\end{figure}
\begin{figure}[H]
  \centering
  \includegraphics[width=0.7\textwidth]{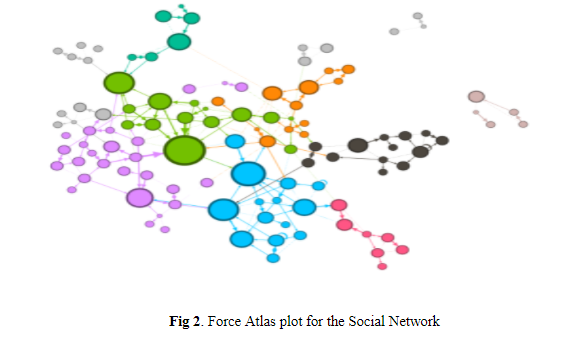}
\end{figure}

\section{Inferences and Analysis}
Various metrics such as Average degree, Average weighted degree, Modularity, HITS, Page Rank, Network diameter, Average Clustering Coefficient were estimated through the Data laboratory. Table 1 illustrates the metric for our network.

\begin{table}[ht]
    \centering
    \caption { Metrics and their Inferences}
    \begin{tabular}{|c|L|L|}
        \hline
        \textbf{Metric} & \textbf {Value} & \textbf{Inferences} \\
        \hline 
        Indegree & Highest - 10 (ID - 039)
     & \multicolumn{1}{m{6cm}|}{Number of students wishing to work with a student. The highest indegree is the most popular student (039).}   \\
        \hline
        Outdegree & Highest - 5  (Various IDs)
     & \multicolumn{1}{m{6cm}|}{Number of students that a candidate is wishing to work with.}  \\
        \hline
        Average Degree of Digraph & 2.097
     & \multicolumn{1}{m{6cm}|}{This implies that every student is connected to at least two students on average.}   \\
        \hline
        Average Degree of Digraph & 2.097
     & \multicolumn{1}{m{6cm}|}{This implies that every student is connected to at least two students on average.}   \\
        \hline
        Network Diameter & 13
     & \multicolumn{1}{m{6cm}|}{Implies the longest path between 2 nodes}   \\
        \hline
        Page Rank & Highest- 0.0478 (ID - 039)
     & \multicolumn{1}{m{6cm}|}{Page Rank of a node conveys the importance of that node in the network, i.e. higher the page rank}   \\
        \hline
        Clustering Coefficient (CC) & Average - 0.189
     & \multicolumn{1}{m{6cm}|}{The possibilities of new relations that can be formed in a network. If a node has lower CC, then there are higher chances of new friendships being formed among the node’s friends}   \\
        \hline
        Connected Components & 3
     & \multicolumn{1}{m{6cm}|}{Implies that two groups are isolated from the more extensive component network, i.e., they do not prefer to do projects outside of their group.}   \\
        \hline
        Modularity Class & 9 Classes
     & \multicolumn{1}{m{6cm}|} {It groups together students who have common interests }   \\
        \hline
    \end{tabular}
    
\end{table}

\subsection{Further Inferences and reasoning}

\begin{figure}
\centering
\begin{minipage}{.55\textwidth}
  \centering
  \includegraphics[width=.6\textwidth]{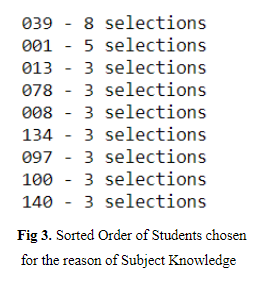}
\end{minipage}%
\begin{minipage}{.55\textwidth}
  \centering
  \includegraphics[width=.6\textwidth]{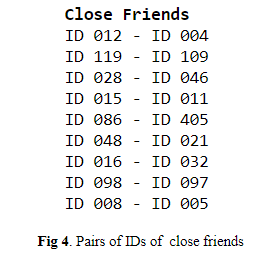}
\end{minipage}
\end{figure}
In particular, analysis was done for the choices of students who entered Subject knowledge as the reason for choosing their teammate.  It was inferred that the student with the higher indegree was chosen most frequently due to their subject knowledge and is the topper (ID - 039). Figure 3 depicts these results.
In \cite{delay2021comparison}, the author analyses how the influence of friends have made a strong impact in forming teams.So an influence parameter is considered to analyse the relative strength in the groups formed.
\newline

2-Member teams lead to dyads, and 3-Member teams lead to triads in the resulting social network. Dyads are only of two types - connected or disconnected . However, triads bring in the concept of structural balance in a graph, which leads to a stable social network only when certain conditions are met. The structural balance concerning this survey is related to how a student chooses their other two teammates. A student cannot just choose two of his close friends; there is also the added requirement that the two know each other for smooth work management.Therefore an analysis was made for mutuality in friendship - pairs who chose each other as their first preference - and the resulting pairs of close friends are listed in Figure 4.

\newpage

\subsection{Types of team formation and algorithms proposed}

 { \bf Ideal Teammate First [ITF]:}
 \newline
 Teams formed from the survey mentioned above, where students were asked to fill teammates they would most likely choose. Hence these teams would be each student’s ideal team without any. Therefore it is expected that each team is more likely to perform well. 
 \newline
 {\bf Algorithm:}
 \newline
 In this algorithm, teams are chosen by students without any real-world constraints as it was not implemented practically. In this sense, they could choose any of the other students who they prefer, regardless of their availability. Hence, the teams chosen will be indefectible and most likely to perform well. Here is the stepwise algorithm of this method and how a social network can be formed: 
 \\
 Constraints: None
 \begin {enumerate}
     \item
  Start 
     \item
  Select a student as node N from the list of respondents. 
     \item
  Populate all the teammate preferences of node N as its neighbors in the social network.
     \item
  Go to the next student in the list and repeat step 3 for the new student. (If any of the neighboring nodes repeat then do not add a new node, instead use the same node with a new connection i.e., strictly one node per student )
     \item
  Repeat steps 3 and 4 until all students’ teammate preferences are accounted for.
     \item
  Stop
 \end{enumerate}
  
    { \bf Ideal Available Teammate First [IATF] :}
    \newline
     Teams taken into consideration were the ones that were formed during an actual project conducted by the department in the 6th semester for a course (SNA) for an elective. Students had the choice to choose their teammates from their section or from a different section. Hence, the teams were expected to closely resemble their ideal choices, however with the exception that some students were most sought after by their peers, and therefore some teams had to change their preferences.
  \newline

    { \bf Team Leader First [TLF]:}
    \newline
     \cite{robert2018you} author Robert states, the more shared leadership a team employs, the more likely its members are to be deriving satisfaction from their cooperation and mutual respect. When this occurs, increases in trust are less likely to correspond directly with increases in satisfaction. Consequently, the relationship between trust and satisfaction weakens the more a virtual team relies on shared leadership. The faculty of the department conducted Codeathon as part of the assessment of a course. Teams were framed by the faculty in charge, wherein the first teammate was annotated to be the leader and best performer. The remaining two members were average performers. Faculty felt the need to know the impact and leadership of the best performer on the team.

\newpage
\begin{table}[ht]
    \centering
    \caption {Statistics of team formation methods}
    \begin{tabular}{|c|L|L|L|}
        \hline
        \textbf{Metric} & \textbf {Ideal Teammate First} & \textbf{Ideal Available Teammate First} & \textbf {Team Leader First}  \\
        \hline
        No. of teams & 75 & 24 & 44 \\
        \hline
        Avg score of students' annotations & 9.06  & 8.75 & 8.18\\
        \hline
        Average marks that students scored   & NA & 7.7 & 6.65 \\
        \hline
        Highest marks scored by a team & NA & 10 & 9 \\
        \hline
        Lowest marks scored by a student & NA & 6 & 3 \\
        \hline
        All teammates alloted same marks? & NA & Yes & No \\
        \hline
        Pros & \multicolumn{1}{m{3.5cm}|}{Teams if formed similarly are most likely to perform well, due to better experience.}  & \multicolumn{1}{m{3.5cm}|}{Most teams are likely to perform well as it was formed out of a shared interest} & \multicolumn{1}{m{3.5cm}|}{Teaches students to handle new teammates of different personalities and varied ideologies.}  \\
        \hline
        Cons & \multicolumn{1}{m{3.5cm}|}{Does not consider the real world constraint of availability of students.}  & \multicolumn{1}{m{3.5cm}|}{Mostly an FCFS model, as some students would hesitate saying no to a student who asks first even though they are not the ideal choice.} & \multicolumn{1}{m{3.5cm}|}{Takes more time for most teams to get along, hence they are less likely to perform to their fullest for the entire duration of the project.}  \\
        \hline
        
    \end{tabular}
    \label{tab:my_label}
\end{table}

\section{Statistical results and their interpretation}

\textbf{1.}  90.7\% of students choosing teammates they have previously worked with implies that most are unwilling to step out of their comfort zones. 
\begin{figure}[H]
  \centering
  \includegraphics[width=0.6\textwidth]{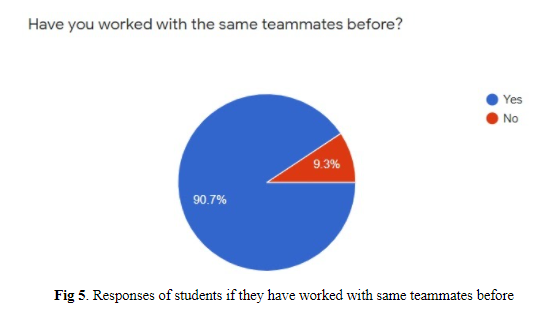}
\end{figure}
   
\textbf{2.}  Almost 90\% - symbol  of students tend to choose their friends as teammates. In comparison, only 48\%- symbol are unwilling students, requiring teammates who know the domain expertise well.
\begin{figure}[H]
  \centering
  \includegraphics[width=0.7\textwidth]{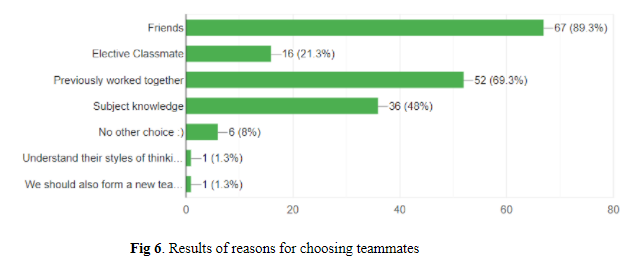}
\end{figure}
    
\textbf{3.}  70\% of students willing to work with other section students, implies the presence of various local bridges. The presence of local bridges is crucial as they are the gateway for new information exchange between classes. 
\begin{figure}[H]
  \centering
  \includegraphics[width=0.6\textwidth]{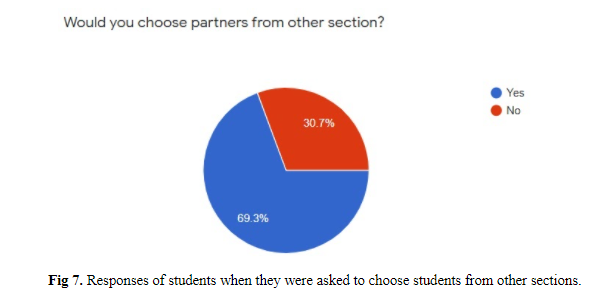}
\end{figure}
    
\textbf{4.}  Surprisingly, 67\% of students prefer the online mode of working together rather than the offline mode. This data visualization implies the impact of the pandemic on students mindsets.

\begin{figure}[H]
  \centering
  \includegraphics[width=0.6\textwidth]{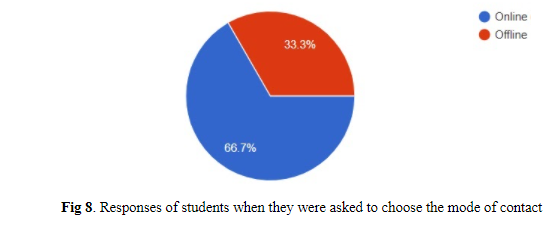}
\end{figure}

\section{Conclusion and Future Scope}
This paper discusses how social network analysis can be applied to study relationships in a college department across various sections. For this, responses were collected from students of all sections in a department asking about their selections for project teammates. This data was then processed and cleaned. A social network was built using the Gephi tool. Various inferences were derived from the metrics as well as the networks obtained.  Various friendship patterns and their subtleties were observed by further analysis through simple python scripts and data analysis techniques. 
Therefore,  Social network analysis of simple data collection among a group of people can corroborate various known inferences and bring new and riveting results.
Further enhancements in this project would be to conduct a similar survey among juniors of the same department and analyze how their responses vary or match. It would be interesting to observe the changes that online learning has catalyzed in students who know each other only through the virtual world.

\bibliographystyle{unsrt}
\bibliography{mybib}

\begin{thebibliography}{10}

\bibitem{oloritun2013identifying}
Rahman~O Oloritun, Anmol Madan, Alex Pentland, and Inas Khayal.
\newblock Identifying close friendships in a sensed social network.
\newblock {\em Procedia-Social and Behavioral Sciences}, 79:18--26, 2013.

\bibitem{saeidibonab2017homophily}
Sepehr Saeidibonab.
\newblock Homophily and friendship dynamics: An analysis of friendship
  formation with respect to homophily principle and distinctiveness theory,
  2017.

\bibitem{noguera2018collaborative}
Ingrid Noguera, Ana-Elena Guerrero-Rold{\'a}n, and Ricard Mas{\'o}.
\newblock Collaborative agile learning in online environments: Strategies for
  improving team regulation and project management.
\newblock {\em Computers \& Education}, 116:110--129, 2018.

\bibitem{grunspan2014understanding}
Daniel~Z Grunspan, Benjamin~L Wiggins, and Steven~M Goodreau.
\newblock Understanding classrooms through social network analysis: A primer
  for social network analysis in education research.
\newblock {\em CBE—Life Sciences Education}, 13(2):167--178, 2014.

\bibitem{pociask2017does}
Sarah Pociask, David Gross, and Mei-Yau Shih.
\newblock Does team formation impact student performance, effort and attitudes
  in a college course employing collaborative learning?.
\newblock {\em Journal of the Scholarship of Teaching and Learning},
  17(3):19--33, 2017.

\bibitem{azizan2018improving}
MT~Azizan, N~Mellon, RM~Ramli, and S~Yusup.
\newblock Improving teamwork skills and enhancing deep learning via development
  of board game using cooperative learning method in reaction engineering
  course.
\newblock {\em Education for Chemical Engineers}, 22:1--13, 2018.

\bibitem{agneessens2022collecting}
Filip Agneessens and Giuseppe~Joe Labianca.
\newblock Collecting survey-based social network information in work
  organizations.
\newblock {\em Social Networks}, 68:31--47, 2022.

\bibitem{gibbs2017investigating}
Jennifer~L Gibbs, Anu Sivunen, and Maggie Boyraz.
\newblock Investigating the impacts of team type and design on virtual team
  processes.
\newblock {\em Human Resource Management Review}, 27(4):590--603, 2017.

\bibitem{delay2021comparison}
Dawn DeLay, Brett Laursen, Noona Kiuru, Adam Rogers, Thomas Kindermann, and
  Jari-Erik Nurmi.
\newblock A comparison of dyadic and social network assessments of peer
  influence.
\newblock {\em International journal of behavioral development},
  45(3):275--288, 2021.

\bibitem{robert2018you}
Lionel~P Robert~Jr and Sangseok You.
\newblock Are you satisfied yet? shared leadership, individual trust, autonomy,
  and satisfaction in virtual teams.
\newblock {\em Journal of the association for information science and
  technology}, 69(4):503--513, 2018.

\end{thebibliography}

\end{document}